\global\def\draftcontrol{0}

   \def\versionno{ sshydro -- draft   }

\catcode`\@=11

\expandafter\ifx\csname draftcontrol\endcsname\relax\global\def\draftcontrol{0}
\fi

{\count255=\time\divide\count255 by 60
\xdef\hourmin{\number\count255}
\multiply\count255 by-60\advance\count255 by\time
\xdef\hourmin{\hourmin:\ifnum\count255<10 0\fi\the\count255}}
\def\draftdate{\number\month/\number\day/\number\year\ \ \ \hourmin }

\newcommand\makepapertitle{\par
  \begingroup
    \renewcommand\thefootnote{\@fnsymbol\c@footnote}%
    \def\@makefnmark{\rlap{\@textsuperscript{\normalfont\@thefnmark}}}%
    \long\def\@makefntext##1{\parindent 1em\noindent
            \hb@xt@1.8em{%
                \hss\@textsuperscript{\normalfont\@thefnmark}}##1}%
     \newpage
     \global\@topnum\z@   
     \@makepapertitle
     \thispagestyle{empty}\@thanks
  \endgroup
  \setcounter{footnote}{0}%
  \global\let\thanks\relax
  \global\let\makepapertitle\relax
  \global\let\@makepapertitle\relax
  \global\let\@thanks\@empty
  \global\let\@author\@empty
  \global\let\@date\@empty
  \global\let\@title\@empty
  \global\let\title\relax
  \global\let\author\relax
  \global\let\date\relax
  \global\let\and\relax
  \def\version{\let\version\@version\@gobble}
}
\def\@makepapertitle{%
  \newpage
   \ifnum\draftcontrol=1 {}
   \version\versionno
   \vskip 3em%
   \else
   \hfill\hbox to 3cm {\parbox{4cm}{\@pubnum}\hss}%
   \vskip 3em%
   \fi
   \begin{center}%
   \let \footnote \thanks
     {\LARGE {\@title}}%
     \vskip 1.5em%
     {\normalsize
       \lineskip .5em%
       \begin{tabular}[t]{c}%
         \@author
       \end{tabular}\par}%
     \vskip 1.5em%
     {\@bstract}%
     \end{center}%
     \vskip 1.5em
     \@date%
   \par
}

\gdef\@pubnum{}
\def\pubnum#1{%
  \gdef\@pubnum{#1}}

\gdef\@bstract{}
\def\Abstract#1{%
  \gdef\@bstract{%
   \parbox{\textwidth-0pc}{%
   \centerline{\bf Abstract}\penalty1000%
\kern.2cm%
\noindent
\renewcommand\baselinestretch{1.0}%
{#1}}}
}

\def\ps@paper{\let\@mkboth\@gobbletwo%
     \ifnum\draftcontrol=1
    \def\@oddfoot{\hbox to \textwidth{\tiny \versionno \hfil\tiny\draftdate}%
    \hskip -\textwidth \hbox to \textwidth{\hfil\rm\thepage\hfil}}%
     \else\def\@oddfoot{\hbox to \textwidth{\hfil\rm\thepage\hfil}}
     \fi
     \let\@evenfoot\@oddfoot
}

\def\body{\clearpage
          \pagestyle{paper}
    }

\def\@version#1{\ifnum\draftcontrol=1
\typeout{}\typeout{#1}\typeout{}
\vskip3mm\centerline{\hbox{\fbox{\normalsize{\tt DRAFT -- #1 -- }
                   {\draftdate}}}}\vskip3mm
\fi}
\let\version\@version
\long\def\eqlabel#1{\ifnum\draftcontrol=1
                    \tag@false  
                    \tag*{(\theequation) \hbox to -0.2cm{\hspace{0cm}\small{#1}\hss}}
                    \refstepcounter{equation}
                    \edef\@currentlabel{\theequation}
                    \ltx@label{#1}          
                    \else
                    \label{#1}
                    \fi
                    }
\let\st@bibitem\@bibitem
\let\st@lbibitem\@lbibitem
\ifnum\draftcontrol=1
  \def\@bibitem#1{%
    \st@bibitem{#1}\a@@label{#1}\ignorespaces}
  \def\@lbibitem[#1]#2{%
    \st@lbibitem[#1]{#2}\a@@label{#2}\ignorespaces}
  \def\a@@label#1{%
    \gdef\a@lab{\smash{\normalfont\small#1}}
    \ifvmode
      \if@inlabel
        \global\setbox\@labels\hbox{%
          \llap{\a@lab\let\a@lab\relax
                \kern\@totalleftmargin\kern\marginparsep}%
          \box\@labels}%
      \fi
    \fi}
\fi

\documentclass[12pt,letterpaper]{article}

\usepackage{amsmath,amssymb,array,calc,epsfig}

\ifnum\draftcontrol=1
\fi

\renewcommand\baselinestretch{1.25}
\renewcommand\section{\@startsection {section}{1}{\z@}%
                                   {-3.5ex \@plus -1ex \@minus -.2ex}%
                                   {2.3ex \@plus.2ex}%
                                   {\normalfont\large\bfseries}}
\renewcommand\subsection{\@startsection{subsection}{2}{\z@}%
                                   {-3.25ex\@plus -1ex \@minus -.2ex}%
                                   {1.5ex \@plus .2ex}%
                                   {\normalfont\normalsize\bfseries}}
\renewcommand\subsubsection{\@startsection{subsubsection}{3}{\z@}%
                                   {-3.25ex\@plus -1ex \@minus -.2ex}%
                                   {1.5ex \@plus .2ex}%
                                   {\normalfont\normalsize\it}}
\renewcommand\paragraph{\@startsection{paragraph}{4}{\z@}%
                                   {-3.25ex\@plus -1ex \@minus -.2ex}%
                                   {1.5ex \@plus .2ex}%
                                   {\normalfont\normalsize\bf}}

\numberwithin{equation}{section}

\def\ie{{\it i.e.}}

\def\revise#1       {\raisebox{-0em}{\rule{3pt}{1em}}%
                     \marginpar{\raisebox{.5em}{\vrule width3pt\
                     \vrule width0pt height 0pt depth0.5em
                     \hbox to 0cm{\hspace{0cm}{%
                     \parbox[t]{4em}{\raggedright\footnotesize{#1}}}\hss}}}}

\def\calc         {{\cal C}}

\def\calf         {{\cal F}}

\def\calk         {{\cal K}}

\def\calm         {{\cal M}}

\def\calp         {{\cal P}}

\def\del          {\partial}

\def\sqr#1#2{{\vcenter{\vbox{\hrule height.#2pt
 \hbox{\vrule width.#2pt height#1pt \kern#1pt
 \vrule width.#2pt}\hrule height.#2pt}}}}

\newcommand{\ft}[2]{{\textstyle{\frac{#1}{#2}}}}

\newcommand{\qq}{\mathfrak{q}}
\newcommand{\ww}{\mathfrak{w}}

\def\w{\omega}

\def\dd{\delta}

\def\hh{\hat{h}}
\def\aa1{\phi}
\def\cc1{\psi}
\def\hh{\hat{h}}

\def\k{\kappa}

\catcode`\@=12

\begin{document}


\title{Hydrodynamics of Sakai-Sugimoto model in the quenched approximation}

\pubnum{%
UWO-TH-06/05}
\date{May 2006}

\author{
Paolo Benincasa$ ^1$ and  Alex Buchel$ ^{1,2}$ \\[0.4cm]
\it $ ^1$Department of Applied Mathematics\\
\it University of Western Ontario\\
\it London, Ontario N6A 5B7, Canada\\[0.2cm]
\it $ ^2$Perimeter Institute for Theoretical Physics\\
\it Waterloo, Ontario N2J 2W9, Canada\\
}

\Abstract{
We study transport properties of the finite temperature Sakai-Sugimoto
model. The model represents a holographic dual to $4+1$ dimensional
supersymmetric $SU(N_c)$ gauge theory compactified on a circle with
anti-periodic boundary conditions for fermions, coupled to $N_f$
left-handed quarks and $N_f$ right-handed quarks localized at
different points on the compact circle. We analytically compute the
speed of sound and the sound wave attenuation in the quenched
approximation. Since confinement/deconfinement (and the chiral
symmetry restoration) phase transitions are first order in this model,
we do not see any signature of these phase transitions in the
transport properties.
 }

\makepapertitle

\body

\version\versionno

\section{Introduction}
Recently Sakai and Sugimoto (SS) \cite{SS,ss2} introduced a supergravity model realizing holographic dual 
\cite{m9711,m2}
to four-dimensional, large $N_c$ QCD with massless flavors. 
Specifically, they considered five-dimensional $SU(N_c)$ maximally supersymmetric 
Yang-Mills (SYM) theory compactified on a circle with anti-periodic boundary conditions for the fermions\footnote{Such boundary 
conditions completely break the supersymmetry and give masses to adjoint fermions of the 5d SYM theory.}, 
and coupled to  $N_f$ left-handed quarks and $N_f$ right-handed 
quarks localized at different points on the compact circle. At weak coupling the model can be represented 
by intersecting $D4/D8/\overline{D8}$ brane system in type IIA 
string theory compactified on a circle. Coincident $N_c$ $D4$ branes wrap the compactification circle, while two stacks 
(with  $N_f$ branes in each) of $D8$ and $\overline{D8}$ branes are localized at different points on this 
compactification circle. At strong coupling, the wrapped $D4$ branes are replaced with an appropriate 
near horizon geometry, while the $D8$ and $\overline{D8}$ branes are treated in the probe 
approximation\footnote{ In other words, the eight-brane backreaction on the $D4$ brane bulk geometry is neglected.}.
The probe brane approximation is valid in the low-energy limit, and as long as $N_f\ll N_c$. The probe approximation 
of the dual holographic description corresponds to a quenched approximation for the fundamental quarks on the 
gauge theory side. 
 
One of the most interesting aspects of the SS model is that it provides a simple holographic realization
of the nonabelian chiral symmetry breaking. In \cite{ASY,ps} the authors studied finite temperature 
confinement/deconfinement and chiral symmetry restoration in SS model. Rather interestingly, these two 
phase transitions are not necessarily simultaneous: for small enough separation of quarks on the circle, 
it was found \cite{ASY} that there is a phase of the hot SS gauge theory which is deconfined, but with a 
broken chiral symmetry. As the separation between quarks exceeds certain critical value, both 
the deconfinement and the chiral symmetry restoration occur at the same temperature.  
For all range of parameters, all of these phase transitions are of first order. Moreover, 
each phase, even being thermodynamically unfavorable, \ie, having a larger free energy, 
appears to exist at arbitrary temperature\footnote{This should be contrasted with a model with 
abelian chiral symmetry breaking \cite{ks}, where the chirally symmetric phase is believed to exist only 
for sufficiently high temperature \cite{kt1,kt2,kt3,kt4} even though the chiral symmetry restoration phase
transition is expected to be  first order.}.

In this paper we study transport properties of the hot SS gauge theory plasma in the quenched approximation. 
Since the backreaction of the $D8$ and $\overline{D8}$ branes is neglected, effectively, we study the 
hydrodynamics of near-extremal $D4$ branes wrapped on a circle with anti-periodic boundary conditions 
for the fermions. The latter model was discussed  in \cite{wth} as the first example of the confining theory 
constructed within gauge theory-string theory correspondence. Since the background geometry 
satisfies condition of \cite{bl,b1}, the shear viscosity $\eta$ of the SS plasma saturates the 
universal viscosity bound proposed\footnote{The universality of the shear viscosity in the supergravity approximation 
was proven in \cite{bl,s1,b1}. } in  \cite{kss} 
\begin{equation}
\frac{\eta}{s}=\frac{1}{4\pi}\,,
\eqlabel{shear}
\end{equation}
where $s$ is the entropy density. On the other hand, the speed of sound and the sound wave attenuation is gauge 
theory specific. Given the dispersion relation for the sound waves
\begin{equation}
\w(q)=v_s q -i\ \frac{2q^2}{3 T }\ \frac{\eta}{s}\ \left(1+\frac{3\zeta}{4\eta}\right)\,, 
\eqlabel{dispertion}
\end{equation} 
where $v_s,  \zeta$ are the plasma sound speed,  
and bulk viscosity correspondingly, for the SS model (in the deconfined phase) we find
\begin{equation}
v_s=\frac{1}{\sqrt{5}}\,,\qquad \frac{\zeta}{\eta}=\frac{4}{15} \,.
\eqlabel{results}
\end{equation}
Notice that transport coefficients \eqref{results} are not dissimilar 
from the  transport properties of other examples of strongly coupled near-conformal  gauge theory plasma \cite{bn1,bn2}
\begin{equation}
\biggl(v_s^2-\frac 13\biggr)\ll 1\,,\qquad 
\frac{\zeta}{\eta}\simeq -\kappa\ \biggl(v_s^2-\frac 13\biggr)\,,\qquad \kappa\sim 1\,,
\eqlabel{nonconsf}
\end{equation}
In fact, the precise value of $\kappa=2$ from \eqref{results} is exactly the same as for the
cascading gauge theory \cite{bn2}. 
Also, there is no signature of the confinement/deconfinement phase 
transition in the sound wave dispersion relation. The latter is not unexpected, given that 
this phase transition is a first order.

The rest of this paper paper is the derivation of \eqref{results}. 
In the next section we discuss effective five-dimensional action  of the near-extremal wrapped 
$D4$ brane system. Interestingly, the resulting five-dimensional supergravity action is 
very similar to the one obtained in \cite{kt4,bn2}. In section 3 we study  
fluctuations of the corresponding black brane geometry dual to a sound 
wave mode of the quenched SS gauge theory plasma. We  introduce gauge invariant 
fluctuations and obtain their equations of motion. These equations of motion are valid beyond the 
hydrodynamic approximation, and for arbitrary temperature. In section 4 we derive and solve fluctuation equations 
analytically in the hydrodynamic limit. 
Imposing Dirichlet condition on the gauge invariant fluctuations at the boundary of the 
background black brane geometry determines \cite{set} 
the dispersion relation for the lowest quasinormal frequency \eqref{dispertion}.
Finally, using the universality result \eqref{shear} we can extract \eqref{results}.

It would be very interesting to extend our computation beyond the probe approximation. We 
expect that the speed of sound waves and their attenuation 
 would develop in this case  dependence on the number of fundamental 
flavors $N_f$, and the dependence on the radius of the  compactification circle.  
Relevant discussion  of the related supergravity background was presented in \cite{leo1}.

\section{Consistent Kaluza-Klein Reduction to $5$-dimensions}

In this section we derive the $5$-dimensional effective action from the type IIA 
$10$-dimensional supergravity action by reduction on $S^{1}\times S^{4}$. 

Consider the type IIA supergravity action in the Einstein frame\footnote{We use conventions  of
\cite{2a} and keep only relevant fields. 
}:
\begin{equation}\label{10IIASugra}
S_{IIA} = \frac{1}{2\kappa_{10}^{2}}
	  \int_{\mathcal{M}_{10}}d^{10}x\:
	   \left(-G^{\mbox{\tiny{$(10)$}}}\right)^{1/2}
	   \left[
	    R^{\mbox{\tiny{$(10)$}}}-\frac{1}{2}\nabla_{\mu}\Phi\nabla^{\mu}\Phi-
	    e^{\Phi/2}|F_{4}|^{2}
	   \right]\,,
\end{equation}
where $\Phi$ is the dilaton and $F_{4}$ is the 4-form field strength, and the 
following metric ansatz:
\begin{equation}\label{10ansatz}
\begin{split}
ds_{10}^{2} & \: = \: G_{MN}^{\mbox{\tiny{$(10)$}}}dx^{M}dx^{N} = \\
            & \: = \: e^{-\frac{10}{3}f}g_{\mu\nu}dx^{\mu}dx^{\nu}+
	              e^{2f}
		      \left[
		       e^{8w}\left(dS^{1}\right)^{2}+e^{-2w}\left(dS^{4}\right)^{2}
		      \right]\, ,
\end{split}
\end{equation}
where the capital Latin indexes ($M,\,N,\,\ldots$) run from $0$ to $9$ and the Greek
indexes ($\mu,\,\nu,\ldots$) run from $0$ to $4$. Moreover, the fields $f$ and $w$ as 
well as $\Phi$ and $F_{4}$ do not depend on the coordinates of $S^{1}\times S^{4}$. 
The 4-form field strength $F_{4}$ is given by
\begin{equation}\label{4form}
F_{4} \: = \: \frac{1}{\sqrt{4!}}A\,\omega_{\mbox{\tiny{$S^{4}$}}} \, ,
\end{equation}
where $A$ is a constant and $\omega_{\mbox{\tiny{$S^{4}$}}}$ is the 4-sphere
volume form.
With such an ansatz, we obtain:
\begin{equation}\label{redrels}
\begin{split}
& \left(-G^{\mbox{\tiny{$(10)$}}}\right)^{1/2} \,=\,
 \left(-g\right)^{1/2}\,e^{-\frac{10}{3}f}\,
 \left(g_{\mbox{\tiny{$4$}}}\right)^{1/2}\,, \\
& \left(-G^{\mbox{\tiny{$(10)$}}}\right)^{1/2}|F_{4}|^{2} \,=\,
 A^{2}\,e^{-8(f-w)}\,\left(-g\right)^{1/2}\,e^{-\frac{10}{3}f}\,
 \left(g_{\mbox{\tiny{$4$}}}\right)^{1/2}\,, \\
& \left(-G^{\mbox{\tiny{$(10)$}}}\right)^{1/2}(\partial_{\mu}\Phi) (\partial^{\mu}\Phi) \,=\, 
 \left(-g\right)^{1/2}\left(g_{\mbox{\tiny{$4$}}}\right)^{1/2}
 (\partial_{\mu}\Phi)(\partial^{\mu}\Phi)\, ,
\end{split}
\end{equation}
where $g_{4}$ is the determinant of the metric of the 4-sphere, and the curvature
scalar $R^{\mbox{\tiny{$(10)$}}}$ is given by:
\begin{equation}\label{R10red}
R^{\mbox{\tiny{$(10)$}}} \,=\, 
  e^{\frac{10}{3}f}
  \left[
   R^{\mbox{\tiny{$(5)$}}}-
   20\,g^{\mu\nu}(\partial_{\mu}w)(\partial_{\nu}w)-
   \frac{40}{3}g^{\mu\nu}(\partial_{\mu}f)(\partial_{\nu}f)
  \right]
 + 12\,e^{-2(f-w)}\, ,
\end{equation}
where $R^{\mbox{\tiny{$(5)$}}}$ is the curvature scalar with respect to the 5-dimensional
metric $g_{\mu\nu}$.
From (\ref{redrels}) and (\ref{R10red}), and integrating over $S^{1}\times S^{4}$ in
the action (\ref{10IIASugra}), the effective 5-dimensional action follows:
\begin{equation}\label{5deffact}
S_5\,=\,\frac{2\pi\,V_{4}}{2\kappa_{10}^2}\int_{\calm_5}\,d^{5}x\,
 \left(-g\right)^{1/2}
 \left[
  R^{\mbox{\tiny{$(5)$}}}-\frac{40}{3}(\del f)^2-20(\del w)^2-\frac 12 (\del\Phi)^2-\calp
 \right]\,,
\end{equation}
where $V_{4}$ is the volume of the 4-sphere and
\begin{equation}\label{potential}
\calp\,\equiv\,
   A^{2}\,e^{\frac{1}{2}\Phi}\,e^{-\frac{34}{3}f+8w}-12\,e^{-\frac{16}{3}f+2w}\, .
\end{equation}
Notice that \eqref{5deffact} is very similar to the five dimensional effective action of the 
cascading gauge theory derived in \cite{kt4,bn2}. 
From \eqref{5deffact} we obtain the following equations of motion:
\begin{equation}
\begin{split}
\Box f-\frac{3}{80}\ \frac{\del\calp}{\del f}\,=\,0\,,
\end{split}
\eqlabel{eqf}
\end{equation}
\begin{equation}
\begin{split}
\Box w-\frac{1}{40}\ \frac{\del\calp}{\del w}\,=\,0\,,
\end{split}
\eqlabel{eqw}
\end{equation}
\begin{equation}
\begin{split}
\Box \Phi- \frac{\del\calp}{\del \Phi}\,=\,0\,,
\end{split}
\eqlabel{eqphi}
\end{equation}
\begin{equation}
\begin{split}
R_{\mu\nu}^{\mbox{\tiny{$(5)$}}}\,=\,&\frac{40}{3}\ \del_\mu f\del_\nu f+20\ \del_\mu w\del_\nu w+\frac 12\  \del_\mu \Phi\del_\nu \Phi+\frac 13 g_{\mu\nu}\ \calp\,.
\end{split}
\eqlabel{eqE}
\end{equation}
Consider now the following ansatz for the 5-dimensional metric:
\begin{equation}
ds_5^2\,=\,-c_1^2\ dt^2+c_2^2\ d\vec{x}^2+c_3^2\ dr^2\,,
\eqlabel{bb0}
\end{equation}
the correspondent 10-dimensional line element therefore takes the form:
\begin{equation}\label{10bb0}
ds_{10}^2\,=\,e^{-\frac{10}{3}f}
 \left[-c_1^2\ dt^2+c_2^2\ d\vec{x}^2+c_3^2\ dr^2\right]
 +\,e^{2f+8w}\left(dS^{1}\right)^{2}\,+\,e^{2(f-w)}\left(dS^{4}\right)^{2}\, .
\end{equation}
A comparison with the finite temperature bulk geometry of the Sakai-Sugimoto model  in the deconfined phase
which is given by \cite{wth,ASY}:
\begin{equation}
\begin{split}
 ds_{\mbox{\tiny{$SS$}}}^{2}\,=&\,
   g_{s}^{-\ft 12}
   \biggl\{
   \left(\frac{r}{R_{D4}}\right)^{\ft 98}
   \left[-\triangle(r)\ dt^{2}+d\vec{x}^2+\left(dS^{1}\right)^{2}\right]\\
&+
   \left(\frac{R_{D4}}{r}\right)^{15/8}
   \left[\frac{dr^{2}}{\triangle(r)}+r^{2}\left(dS^{4}\right)^{2}\right]
   \biggr\}\,,\qquad  F_{4}\,=\,\frac{2\pi N_{c}}{V_{4}}\omega_{\mbox{\tiny{$S^{4}$}}}\, ,\\
&
  e^{\Phi}\,=\,g_{s}\left(\frac{r}{R_{D4}}\right)^{3/4}\, ,\quad
  R_{D4}^{3}\,\equiv \,\pi g_{s}N_{c}l_{s}^{3}\, ,\quad
  \triangle(r)\,\equiv\,1-\left(\frac{r_{\Lambda}}{r}\right)^{3}\, ,
\end{split}
\eqlabel{SSbkg}
\end{equation}
leads to the following identifications:
\begin{equation}
\begin{split}
 w &\,=\,\frac{1}{10}\ \ln{r}-\frac{3}{10}\ \ln{R_{D4}}\, , \\
 f &\,=\,-\frac{1}{4}\ \ln{g_{s}}+\frac{51}{80}\ \ln{R_{D4}}+\frac{13}{80}\ \ln{r} \, , \\
 c_{1} &\,=\,g_{s}^{-\ft {2}{3}}\ R_{D4}^{\ft 12}\ r^{\ft 56}\ {[\triangle(r)]}^{\ft 12}\, , \\
 c_{2} &\,=\,c_{1}\ {[\triangle(r)]}^{-\ft 12}\, , \\
 c_{3} &\,=\,{g_{s}^{-\ft {2}{3}}\ R_{D4}^{2}}\ {r^{-\ft 23}\ {[\triangle(r)]}^{-\ft 12}}.
\end{split}
\eqlabel{ssfields}
\end{equation}
In what follows we set $g_{s}=1$ and $R_{D4}=1$ (in this case $A^2=\ft 92$), so that the relations (\ref{ssfields})
are:
\begin{equation}
\begin{split}
 w &\,=\,\frac{1}{10}\ln{r}\, , \\
 f &\,=\,\frac{13}{80}\ln{r}\, , \\
 c_{1} &\,=\,r^{\ft 56}\ {[\triangle(r)]}^{\ft 12}\, , \\
 c_{2} &\,=\,r^{\ft 56}\, , \\
 c_{3} &\,=\,r^{-\ft 23}\ {[\triangle(r)]}^{-\ft 12}.
\end{split}
\eqlabel{ssfields1}
\end{equation}

We conclude this section with short comments on the thermodynamics of the black brane configuration \eqref{SSbkg}.
The Hawking temperature $T$ is related to the nonextremality parameter $r_{\Lambda}$ as  
\begin{equation}
\frac{1}{T}=\frac{4\pi}{3 r_{\Lambda}^{1/2}}
\eqlabel{temperature}
\end{equation}
From \eqref{SSbkg} the entropy density $s$ of the black branes is 
\begin{equation}
s\,\propto\, r_{\Lambda}^{5/2}\, \propto\, T^5
\eqlabel{entdens}
\end{equation}
The first law of thermodynamics 
\[
-dP=dF=-s\ dT\,,
\]
where $P$ is the pressure and $F$ is the free energy density,   
 then implies that 
\begin{equation}
-P=F=-\frac 16\ s T\,\Rightarrow\, \epsilon=\frac 56\ sT\,\Rightarrow\, P=\frac 15\ \epsilon\,,
\eqlabel{eos}
\end{equation}
where $\epsilon$ is the energy density. 
From the equation of state \eqref{eos} we find that 
\begin{equation}
v_s^2=\frac{\del P}{\del \epsilon}=\frac 15\,.
\eqlabel{vseos}
\end{equation}
In the next section we reproduce \eqref{vseos} from the dispersion relation for the 
pole in stress-energy tensor 
two point correlation function in the sound wave channel, or equivalently \cite{set}, 
from the dispersion relation for the lowest sound channel quasinormal mode in the non-extremal geometry 
\eqref{SSbkg}.

\section{Fluctuations}
Now we study  fluctuations in the background geometry
\begin{equation}
\begin{split}
g_{\mu\nu}&\,\to\, g_{\mu\nu}+h_{\mu\nu}\,,\\
f&\,\to\, f+\dd f\,,\\
w&\,\to\, w+\dd w\,,\\
\Phi&\,\to\, \Phi+\dd \Phi\,,
\end{split}
\eqlabel{fluctuations}
\end{equation}
where $\{g_{\mu\nu},f,w,\Phi\}$ are the black brane 
background configuration (satisfying  \eqref{ssfields1}),
and $\{h_{\mu\nu},\dd f,\dd w,\dd \Phi\}$ are the fluctuations. We choose the gauge 
\begin{equation}
h_{tr}=h_{x_ir}=h_{rr}=0\,.
\eqlabel{gaugec}
\end{equation}
 Additionally, 
we assume that all the fluctuations depend only on $(t,x_3,r)$,\ \ie, we have an $O(2)$ rotational symmetry in the 
$x_1x_2$ plane.

At a linearized level we find that the following sets of fluctuations decouple from each other
\begin{equation}
\begin{split}
&\{h_{x_1x_2}\}\,,\\
&\{h_{x_1x_1}-h_{x_2x_2}\}\,,\\
&\{h_{tx_1},\ h_{x_1x_3}\}\,,\\
&\{h_{tx_2},\ h_{x_2x_3}\}\,,\\
&\{h_{tt},\ h_{aa}\equiv h_{x_1x_1}+h_{x_2x_2},\ h_{tx_3},\ h_{x_3x_3},\ \dd f,\ 
 \dd w,\ \dd \Phi\}\,.
\end{split}
\end{equation}
The last set of fluctuations is a  holographic dual to the sound waves in quenched SS gauge theory plasma 
which is of interest here. Introduce
\begin{equation}
\begin{split}
h_{tt}=&c_1^2\ \hh_{tt}=e^{-i\w t+iq x_3}\ c_1^2\  H_{tt}\,,\\
h_{tz}=&c_2^2\ \hh_{tz}=e^{-i\w t+iq x_3}\ c_2^2\  H_{tz}\,,\\
h_{aa}=&c_2^2\ \hh_{aa}=e^{-i\w t+iq x_3}\ c_2^2\  H_{aa}\,,\\
h_{zz}=&c_2^2\ \hh_{zz}=e^{-i\w t+iq x_3}\ c_2^2\  H_{zz}\,,\\
\dd f=&e^{-i\w t+iq x_3}\ \calf\,,\\
\dd w=&e^{-i\w t+iq x_3}\ \Omega\,,\\
\dd \Phi=&e^{-i\w t+iq x_3}\ p\,,\\
\hh_{ii}=&\hh_{aa}+\hh_{zz}\,,\qquad H_{ii}=H_{aa}+H_{zz}\,,
\end{split}
\eqlabel{rescale}
\end{equation} 
where $\{H_{tt},H_{tz},H_{aa},H_{zz},\calf,\Omega,p\}$ are functions of a radial coordinate  only. 
Expanding at a linearized level Eqs.~\eqref{eqf}-\eqref{eqE} 
with Eq.~\eqref{fluctuations} and Eq.~\eqref{rescale} we find
the following coupled system of ODE's
\begin{equation}
\begin{split}
0=&H_{tt}''+H_{tt}'\ \left[\ln\frac{c_1^2c_2^3}{c_3}\right]'-H_{ii}'\ [\ln c_1]'
-\frac{c_3^2}{c_1^2}\left(q^2\frac{c_1^2}{c_2^2}\ H_{tt}+\w^2\ H_{ii}+2\w q\ H_{tz}\right)\\
&-\frac 23 c_3^2 \left(\frac{\del\calp}{\del f}\ \calf+\frac{\del\calp}{\del w}\ \Omega+\frac{\del\calp}{\del \Phi}\ p\right)\,,
\end{split}
\eqlabel{fl1}
\end{equation}
\begin{equation}
\begin{split}
0=&H_{tz}''+H_{tz}'\ \left[\ln\frac{c_2^5}{c_1c_3}\right]'
+\frac{c_3^2}{c_2^2}\ \w q\ H_{aa}\,,
\end{split}
\eqlabel{fl2}
\end{equation}
\begin{equation}
\begin{split}
0=&H_{aa}''+H_{aa}'\ \left[\ln\frac{c_1c_2^5}{c_3}\right]'+(H_{zz}'-H_{tt}')\ [\ln c_2^2]'
+\frac{c_3^2}{c_1^2}\left(\w^2-q^2\frac{c_1^2}{c_2^2}\right)\ H_{aa}\\
&+\frac 43 c_3^2 \left(\frac{\del\calp}{\del f}\ \calf+\frac{\del\calp}{\del w}\ \Omega+\frac{\del\calp}{\del \Phi}\ p\right)\,,
\end{split}
\eqlabel{fl3}
\end{equation}
\begin{equation}
\begin{split}
0=&H_{zz}''+H_{zz}'\ \left[\ln\frac{c_1c_2^4}{c_3}\right]'+(H_{aa}'-H_{tt}')\ [\ln c_2]'\\
&+\frac{c_3^2}{c_1^2}\left(\w^2\ H_{zz}+2\w q\ H_{tz}+q^2\frac{c_1^2}{c_2^2}(H_{tt}-H_{aa})\right)\\
&+\frac 23 c_3^2 \left(\frac{\del\calp}{\del f}\ \calf+\frac{\del\calp}{\del w}\ \Omega+\frac{\del\calp}{\del \Phi}\ p\right)\,,
\end{split}
\eqlabel{fl4}
\end{equation}
\begin{equation}
\begin{split}
0=&\calf''+\calf'\ \left[\ln\frac{c_1c_2^3}{c_3}\right]'+\frac 12 f'\ [H_{ii}-H_{tt}]'+\frac{c_3^2}{c_1^2}
\left(\w^2-q^2\frac{c_1^2}{c_2^2}\right)\ \calf\\
&-\frac {3}{80} c_3^2 \biggl(\frac{\del^2\calp}{\del f^2}\ \calf+\frac{\del^2\calp}{\del f\del w}\ \Omega
+\frac{\del^2\calp}{\del f\del \Phi}\ p\biggr)\,,
\end{split}
\eqlabel{fl5}
\end{equation}
\begin{equation}
\begin{split}
0=&\Omega''+\omega'\ \left[\ln\frac{c_1c_2^3}{c_3}\right]'+\frac 12 w'\ [H_{ii}-H_{tt}]'+\frac{c_3^2}{c_1^2}
\left(\w^2-q^2\frac{c_1^2}{c_2^2}\right)\ \Omega\\
&-\frac {1}{40} c_3^2 \biggl(\frac{\del^2\calp}{\del w\del f}\ \calf+\frac{\del^2\calp}{\del w^2}\ \Omega
+\frac{\del^2\calp}{\del w\del \Phi}\ p\biggr)\,,
\end{split}
\eqlabel{fl6}
\end{equation}
\begin{equation}
\begin{split}
0=&p''+p'\ \left[\ln\frac{c_1c_2^3}{c_3}\right]'+\frac 12 \Phi'\ [H_{ii}-H_{tt}]'+\frac{c_3^2}{c_1^2}
\left(\w^2-q^2\frac{c_1^2}{c_2^2}\right)\ p\\
&-c_3^2 \biggl(\frac{\del^2\calp}{\del \Phi\del f}\ \calf+\frac{\del^2\calp}{\del\Phi\del w}\ \Omega
+\frac{\del^2\calp}{\del \Phi^2}\ p\biggr)\,,
\end{split}
\eqlabel{fl7}
\end{equation}
where all derivatives $\del\calp$ are evaluated on the background geometry. 
Additionally, there are three first order constraints associated with the (partially) fixed diffeomorphism invariance 
\begin{equation}
\begin{split}
0=&\w\left(H_{ii}'+\left[\ln\frac{c_2}{c_1}\right]'\ H_{ii}\right)+q\left(H_{tz}'+2\left[\ln\frac{c_2}{c_1}\right]'\ H_{tz}\right)
\\
&+\w\ \left(\frac{80}{3}f'\calf+40w'\Omega+\Phi'p\right)\,,
\end{split}
\eqlabel{const1}
\end{equation}
\begin{equation}
\begin{split}
0=&q\left(H_{tt}'-\left[\ln\frac{c_2}{c_1}\right]'\ H_{tt}\right)+\frac{c_2^2}{c_1^2}\w\ H_{tz}'-q\ H_{aa}
-q\ \left(\frac{80}{3}f'\calf+40w'\Omega+\Phi'p\right)\,,
\end{split}
\eqlabel{const2}
\end{equation}
\begin{equation}
\begin{split}
0=&[\ln c_1c_2^2]'H_{ii}'-[\ln{c_2^3}]'\ H_{tt}'+\frac{c_3^2}{c_1^2}
\left(\w^2\ H_{ii}+2\w q\ H_{tz}+q^2\ \frac{c_1^2}{c_2^2}\left(H_{tt}-H_{aa}\right)\right)\\
&+c_3^2\left(\frac{\del\calp}{\del f}\ \calf+\frac{\del\calp}{\del w}\ \Omega+\frac{\del\calp}{\del \Phi}\ p\right)
-\left(\frac{80}{3}f'\calf'+40w'\Omega'+\Phi'p'\right)\,.
\end{split}
\eqlabel{const3}
\end{equation}
We explicitly verified that Eqs.~\eqref{fl1}-\eqref{fl7} are consistent with constraints 
\eqref{const1}-\eqref{const3}. 

Introducing the gauge invariant fluctuations
\begin{equation}
\begin{split}
Z_H=&4\frac{q}{\w} \ H_{tz}+2\ H_{zz}-H_{aa}\left(1-\frac{q^2}{\w^2}\frac{c_1'c_1}{c_2'c_2}\right)+2\frac{q^2}{\w^2}
\frac{c_1^2}{c_2^2}\ H_{tt}\,,\\
Z_f=&\calf-\frac{f'}{[\ln c_2^4]'}\ H_{aa}\,,\\
Z_w=&\Omega-\frac{w'}{[\ln c_2^4]'}\ H_{aa}\,,\\
Z_\Phi=&p-\frac{\Phi'}{[\ln c_2^4]'}\ H_{aa}\,,
\end{split}
\eqlabel{physical}
\end{equation}
and a new radial coordinate
\begin{equation}
x\equiv \frac{c_1}{c_2}\,,
\eqlabel{defx}
\end{equation}
we find from Eqs.~\eqref{fl1}-\eqref{fl7}, \eqref{const1}-\eqref{const3}, decoupled set of 
equations 
of motion for $Z$'s\, :
\begin{equation}
\begin{split}
0=&{\frac {d^{2}{  Z_H}  }{d{x}^{2}}}+{\frac { \left( 3
\,{q}^{2} \left( 2\,{x}^{2}-1 \right) +5\,{\omega}^{2} \right)  }{x \left( 5\,{\omega}^{2}-{q}^{2}
 \left( 3+2\,{x}^{2} \right)  \right) }}\ {\frac 
{d{  Z_H}}{dx}}\\
&+\frac 49\,{\frac { \left(  \left( 
-{\omega}^{2}+{q}^{2}{x}^{2} \right)  \left( {q}^{2} \left( 3+2\,{x}^{
2} \right) -5\,{\omega}^{2} \right) -18\,{q}^{2}{  r_{\Lambda}}\,{x}^{2}
 \left( 1-{x}^{2} \right) ^{5/3} \right)  }{
 \left( 5\,{\omega}^{2}-{q}^{2} \left( 3+2\,{x}^{2} \right)  \right) 
 \left( 1-{x}^{2} \right) ^{5/3}{x}^{2}{  r_{\Lambda}}}}\ {  Z_H} \\
&+{\frac {4}{15}}\,{
\frac {{q}^{2} \left( -3\,{q}^{2}+5\,{\omega}^{2} \right)   }{{\omega}^{2} \left( 5\,{\omega}^{2}-{q}^
{2} \left( 3+2\,{x}^{2} \right)  \right) }}\ \left( 48\,
{  Z_w}  +9\,{  Z_{\Phi}}  +52\,{  Z_f
}   \right)\,,
\end{split}
\eqlabel{zH}
\end{equation}
\begin{equation}
\begin{split}
0=&{\frac {d^{2}{  Z_f}}{d{x}^{2}}} +{\frac  1{x}}\ \frac {dZ_f}
{dx}
-{\frac {4}{225}}\,{\frac {
 \left( 25\, \left( 1-{x}^{2} \right)  \left( -{\omega}^{2}+{q}^{2}{x}
^{2} \right) +243\,{  r_{\Lambda}}\,{x}^{2} \left( 1-{x}^{2} \right) ^{2/3}
 \right) }{{  r_{\Lambda}}\,{x}^{2} \left( 1-{x}^{2
} \right) ^{8/3}}}\  {  Z_f} \\
&+{\frac {9}{ 25\left( 1-{x}^{2} \right) ^{
2}}}\ \left({  Z_{\Phi}}  +12\,{  Z_w}  \right)\,,
\end{split}
\eqlabel{zf}
\end{equation}
\begin{equation}
\begin{split}
0=&{\frac {d^{2}{  Z_w}}{d{x}^{2}}}  +\frac 1x\  {{\frac {d{  Z_w}}
{dx}}}-{\frac {4}{225}}\,{\frac {
 \left( 25\, \left( 1-{x}^{2} \right)  \left( -{\omega}^{2}+{q}^{2}{x}
^{2} \right) +162\,{  r_{\Lambda}}\,{x}^{2} \left( 1-{x}^{2} \right) ^{2/3}
 \right)  }{{  r_{\Lambda}}\,{x}^{2} \left( 1-{x}^{2
} \right) ^{8/3}}}\ {  Z_w}\\
&+{\frac { 6}{25 \left( 1-{x}^{2} \right) ^{2}
}}\ \left(12\,{  Z_f}  -{  Z_{\Phi}} \right)\,,
\end{split}
\eqlabel{zw}
\end{equation}
\begin{equation}
\begin{split}
0=&{\frac {d^{2}{  Z_{\Phi}}}{d{x}^{2}}}  +\frac 1x\  {{\frac {
d{  Z_{\Phi}}}{dx}}}-{\frac {4}{45}}\,{\frac {
 \left( 5\, \left( 1-{x}^{2} \right)  \left( -{\omega}^{2}+{q}^{2}{x}^
{2} \right) +9\,{  r_{\Lambda}}\,{x}^{2} \left( 1-{x}^{2} \right) ^{2/3}
 \right)}{{  r_{\Lambda}}\,{x}^{2} \left( 1-{x}^
{2} \right) ^{8/3}}}\  {  Z_{\Phi}}\\
&-{\frac {48 }{5 \left( 1-{x}^{2} \right) ^{2}}}\ \left(  {Z_w} -{  Z_f} \right)\,.
\end{split}
\eqlabel{zphi}
\end{equation}
Further decoupling occurs if we introduce 
\begin{equation}
\k\equiv 48\,
{  Z_w}  +9\,{  Z_{\Phi}}  +52\,{  Z_f
}\,.
\eqlabel{defkappa}
\end{equation}
In this case we find\footnote{This set of gauge invariant fluctuations will be sufficient to 
determine the sound wave dispersion relation.}: 
\begin{equation}
\begin{split}
0=&{\frac {d^{2}{  Z_H}  }{d{x}^{2}}}+{\frac { \left( 3
\,{q}^{2} \left( 2\,{x}^{2}-1 \right) +5\,{\omega}^{2} \right)  }{x \left( 5\,{\omega}^{2}-{q}^{2}
 \left( 3+2\,{x}^{2} \right)  \right) }}\ {\frac 
{d{  Z_H}}{dx}}\\
&+\frac 49\,{\frac { \left(  \left( 
-{\omega}^{2}+{q}^{2}{x}^{2} \right)  \left( {q}^{2} \left( 3+2\,{x}^{
2} \right) -5\,{\omega}^{2} \right) -18\,{q}^{2}{  r_{\Lambda}}\,{x}^{2}
 \left( 1-{x}^{2} \right) ^{5/3} \right)  }{
 \left( 5\,{\omega}^{2}-{q}^{2} \left( 3+2\,{x}^{2} \right)  \right) 
 \left( 1-{x}^{2} \right) ^{5/3}{x}^{2}{  r_{\Lambda}}}}\ {  Z_H} \\
&+{\frac {4}{15}}\,{
\frac {{q}^{2} \left( -3\,{q}^{2}+5\,{\omega}^{2} \right)   }{{\omega}^{2} \left( 5\,{\omega}^{2}-{q}^
{2} \left( 3+2\,{x}^{2} \right)  \right) }}\ \k\,,
\end{split}
\eqlabel{zH1}
\end{equation}
\begin{equation}
\begin{split}
0=&{\frac {d^{2}\k}{d{x}^{2}}} +{\frac 1x}\ {{\frac {d\k}{d
x}} }+{\frac {
 4\left( {\omega}^{2}-{q}^{2}{x}^{2} \right) }{{  9 r_{\Lambda}}\,{x}^{2}
 \left( 1-{x}^{2} \right) ^{5/3}}}\ \k\,.
\end{split}
\eqlabel{kappae}
\end{equation}

\section{Hydrodynamic limit}
We study now physical fluctuation equations \eqref{zH1}, \eqref{kappae} 
in the hydrodynamics approximation, $\w\to 0,\ q\to 0$ with $\ft \w q$ kept constant. Similar to the  computations in \cite{bn1,bn2}, 
we would need only leading and next-to-leading (in $q$) solution of \eqref{zH1} and \eqref{kappae}.  
We find that at the  horizon, $x\to 0_+$, $Z_H\propto x^{\pm i\w/(2\pi T)}$, 
and similarly for $\k$. 
Incoming boundary conditions on all physical modes implies that 
\begin{equation}
\begin{split}
&Z_H(x)=x^{-i\ww} z_H(x)\,,\qquad \k(x)=x^{-i\ww} \calk(x)\,,
\end{split}
\eqlabel{incoming}
\end{equation}
where $\{z_H, \calk\}$ are regular at the horizon; we further introduced 
\begin{equation}
\ww\equiv \frac{\w}{2\pi T}\,,\qquad \qq\equiv\frac{q}{2\pi T} \,.
\eqlabel{defwwqq}
\end{equation}
There is a single integration constant for  these  physical modes, namely, 
the overall scale. Without the loss of generality the latter can be fixed as 
\begin{equation}
z_H(x)\bigg|_{x\to 0_+}=1\,.
\eqlabel{bconditions}
\end{equation}
In this case, the pole dispersion relation is simply determined as \cite{set}
\begin{equation}
z_H(x)\bigg|_{x\to 1_-}=0\,.
\eqlabel{poledisp}
\end{equation}
The other boundary condition (besides regularity at the horizon and \eqref{poledisp})
is \cite{set}
\begin{equation}
\calk(x)\bigg|_{x\to 1_-}=0\,.
\eqlabel{rembound}
\end{equation}
Let's  introduce 
\begin{equation}
\begin{split}
&z_H=z_{H,0}+i\ \qq\  z_{H,1}\,,\qquad \calk=\calk_0+i\ \qq\ \calk_1\,,
\end{split}
\eqlabel{defzz}
\end{equation}
where the  index refers to either the leading, $\propto \qq^0$, or to the next-to-leading, $\propto \qq^1$,  
order in the hydrodynamic approximation.  
Additionally, as we are interested in the hydrodynamic pole dispersion relation in the stress-energy correlation functions,
we find it convenient to parameterize 
\begin{equation}
\ww=v_s\ \qq- i\ \qq^2\ \Gamma\,,
\eqlabel{disprel}
\end{equation}
where the speed of sound $v_s$ and the sound wave attenuation $\Gamma$ 
are to be determined from the pole dispersion relation
\eqref{poledisp}
\begin{equation}
\begin{split}
&z_{H,0}\bigg|_{x\to 1_-}=0\,,\qquad z_{H,1}\bigg|_{x\to 1_-}=0\,.
\end{split}
\eqlabel{poledisp1}
\end{equation}
Using  parameterizations 
\eqref{defzz}, \eqref{disprel}, we obtain
from \eqref{zH1} and \eqref{kappae} the following ODE's 
\begin{equation}
\begin{split}
0=&x\ \calk_0''+\calk_0'\,,
\end{split}
\eqlabel{fin1}
\end{equation}
\begin{equation}
\begin{split}
0=&{  z_{H,0}}''  -{\frac { \left( 6
\,{x}^{2}-3+5\,{{  v_s}}^{2} \right)  }{x \left( 2\,{x}^{2}-5\,{{  v_s}}^{2}+3 \right) }}\   z_{H,0}'+{
\frac {8  }{2\,{x}^{2}-5\,{{  v_s}}^{2}+3}}\ {  z_{H,0}}-{\frac { 4\left( -3+5\,{{  v_s}}^{2} \right) {  
\calk_0} }{ 15\left( 2\,{x}^{2}-5\,{{  v_s}}^{2}+3
 \right) {{  v_s}}^{2}}}\,,
\end{split}
\eqlabel{fin2}
\end{equation}
describing leading ($\propto \qq^0$), and 
\begin{equation}
\begin{split}
0=&x\ \calk_1''+\calk_1'-2 v_s\ \calk_0'\,,
\end{split}
\eqlabel{fin3}
\end{equation}
\begin{equation}
\begin{split}
0=&{  z_{H,1}}''  -{\frac { \left( 6
\,{x}^{2}-3+5\,{{  v_s}}^{2} \right)  }{x \left( 2\,{x}^{2}-5\,{{  v_s}}^{2}+3 \right) }}\   z_{H,1}'+{
\frac {8  }{2\,{x}^{2}-5\,{{  v_s}}^{2}+3}}\ {  z_{H,1}}\\
&+
{\frac {2{  v_s}\, \left( 40\,{x}^{2}{  \Gamma}+20\,{x}^{2}{{  v_s
}}^{2}-25\,{{  v_s}}^{4}+30\,{{  v_s}}^{2}-4\,{x}^{4}-12\,{x}^{2}-9
 \right)  }{x \left( 2\,{x}^{
2}-5\,{{  v_s}}^{2}+3 \right) ^{2}}}\ z_{H,0}'\\
&-{\frac {8{  v_s}\, \left( -2
\,{x}^{2}+5\,{{  v_s}}^{2}-3+10\,{  \Gamma} \right) 
 }{ \left( 2\,{x}^{2}-5\,{{  v_s}}^{2}+3 \right) ^{2
}}}\ {  z_{H,0}}+{\frac {8}{15}}\,{\frac {{  \Gamma}\, \left( 6\,{x}^{2}+9+25\,{{
  v_s}}^{4}-30\,{{  v_s}}^{2} \right) {  \calk_0}  }
{ \left( 2\,{x}^{2}-5\,{{  v_s}}^{2}+3 \right) ^{2}{{  v_s}}^{3}}}\\
&-{
\frac {4}{15}}\,{\frac {    \left( -3+5\,{{
  v_s}}^{2} \right)\calk_1 }{ \left( 2\,{x}^{2}-5\,{{  v_s}}^{2}+3 \right) 
{{  v_s}}^{2}}}\,,
\end{split}
\eqlabel{fin4}
\end{equation}
describing next-to-leading ($\propto \qq^1$) order in the hydrodynamic approximation.  
Solving \eqref{fin1} and \eqref{fin3} subject to  regularity at the horizon and the boundary condition  
\eqref{rembound} we find
\begin{equation}
\calk_0=0\,,\qquad \calk_1=0\,.
\eqlabel{ksol}
\end{equation}
Given \eqref{ksol}, solution to \eqref{fin1} subject to regularity at the horizon and the boundary condition
\eqref{bconditions} is 
\begin{equation}
z_{H,0}={\frac {5\,{v_s}^{2}+2\,{x}^{2}-3}{5\,{v_s}^{2}-3}}\,,
\eqlabel{zh0s}
\end{equation}
which from \eqref{poledisp1} determines (in agreement with \eqref{vseos})
\begin{equation}
v_s=\frac{1}{\sqrt{5}}\,.
\eqlabel{vsres}
\end{equation}
With \eqref{ksol}-\eqref{vsres}, solution to \eqref{fin4} subject to regularity at the horizon 
 is
\begin{equation}
z_{H,1}=\calc\ (1-x^2)+\frac{1}{\sqrt{5}}\ \left(5\Gamma-2\right)\,,
\eqlabel{zh1s}
\end{equation}
where $\calc$ is an arbitrary integration constant. From \eqref{poledisp1} we conclude\footnote{Boundary condition 
\eqref{bconditions} fixes $\calc=0$.}
\begin{equation}
\Gamma=\frac 25\,.
\eqlabel{gammas}
\end{equation}
Finally, comparing \eqref{dispertion} and \eqref{disprel}, and using \eqref{shear}, we obtain from \eqref{gammas} 
result quoted in \eqref{results}.

\section*{Acknowledgments}
We would like to thanks Colin Denniston, Martin Mueser and Andrei Starinets 
for valuable discussions. Research at
Perimeter Institute is supported in part by funds from NSERC of
Canada. AB gratefully   acknowledge  support by  NSERC Discovery
grant.

\end{document}